\documentclass{iaus}
\usepackage{graphics}

  \checkfont{eurm10}
  \iffontfound
    \IfFileExists{upmath.sty}
      {\typeout{^^JFound AMS Euler Roman fonts on the system,
                   using the 'upmath' package.^^J}%
       \usepackage{upmath}}
      {\typeout{^^JFound AMS Euler Roman fonts on the system, but you
                   dont seem to have the}%
       \typeout{'upmath' package installed. iaus.cls can take advantage
                 of these fonts,^^Jif you use 'upmath' package.^^J}%
      }
  \else
  \fi

  \checkfont{msam10}
  \iffontfound
    \IfFileExists{amssymb.sty}
      {\typeout{^^JFound AMS Symbol fonts on the system, using the
                'amssymb' package.^^J}%
       \usepackage{amssymb}%

      }{}
  \fi

  \IfFileExists{amsbsy.sty}
    {\typeout{^^JFound the 'amsbsy' package on the system, using it.^^J}%
     \usepackage{amsbsy}}
    {\providecommand\boldsymbol[1]{\mbox{\boldmath $##1$}}}

\newsavebox{\astrutbox}
\sbox{\astrutbox}{\rule[-5pt]{0pt}{20pt}}

\title[Outskirts of Galaxy Clusters: intense life in the suburbs]
      {Application of the Voronoi Tessellation Technique
for Galaxy Cluster Search in the M\"{u}nster Red Sky Survey}

\author[E. Panko {\it et al.\/}]%
{E. Panko$^1$%
\and P. Flin$^2$}

\affiliation{$^1$Astronomical Observatory of the Nikolaev State
University, Nikolaev, Ukraine\\[\affilskip]
$^2$Institute of Physics, Pedagogical University, Kielce, Poland}

\pubyear{2004}
\volume{195}
\pagerange{1--8}
\date{?? and in revised form ??}
\jname{Outskirts of Galaxy Clusters: intense life in the suburbs}
\editors{A. Diaferio, ed.}
\begin{document}

\maketitle

\begin{abstract}
We present the preliminary result of our project, consisting in
studying the properties of a large sample of galaxy clusters. The
M\"{u}nster Red Sky Survey, which is a large galaxy catalogue
covering an area of about 5000 square degrees on the southern
hemisphere serves as our observational basis. It is complete up to
$r_F=18^m.3$. Creation of a cluster catalogue is the first step of
our investigation. We propose to use the 2D Voronoi tessellation
technique for identifying galaxy clusters in this 2D catalogue.
Points with high values of the inverse Voronoi tessel area will be
regarded as galaxy cluster centroids. We show that this approach
works correctly.
\end{abstract}

\section{Introduction}
In this paper we present the first element of our project. It
consists in studying the properties of a large sample of galaxy
clusters. In order to perform such studies, we need a sample of
clusters extracted in a uniform manner from a homogeneous set of
data. Therefore, we chose the M\"{u}nster Red Sky Survey as our
observational basis. The first step of investigation is to create
the catalogue of galaxy clusters. There are three basic cluster
detection algorithms: the matched filter algorithm (Postman et al.
1996), the adaptive matched filter algorithm (Kepner et al. 1999)
and the Voronoi tessellation technique (Icke \& van de Weygaert
1987, Zaninetti 1989, Ramella et al. 1999, 2001). Kim et al.
(2002) made a comparison of these cluster-finding algorithms,
using a Monte Carlo experiment with simulated clusters. We decided
to apply the Voronoi tessellation technique for cluster detection.
The Voronoi tessellation technique is completely non-parametric,
and therefore sensitive to both symmetric and elongated clusters,
allowing correct studies of non-spherically symmetric structures.
For a distribution of seeds, the Voronoi tessellation creates
polygonal cells containing one seed each and enclosing the whole
area closest to its seed. This is the definition of a Voronoi cell
in 2D. This natural partitioning of space by the Voronoi
tessellation has been used to model the large-scale distribution
of galaxies.

\section{Observational data}
The M\"{u}nster Red Sky Survey constitutes the observational basis
of our work. It contains scans of 217 adjoining plates of the ESO
Southern Sky Atlas R covering more than 5000 degrees squared.
Plates were scanned using two PDS 2020 $GM^{plus}$
microdensitometers of the Astronomical Institute in M\"{u}nster.
The star/galaxy classification was performed by an automatic
procedure using effective radius, central intensity and surface
brightness profiles. After that, visual inspection of over 2 775
000 objects was performed, allowing their re-classification.
Uniformity of the whole catalogue was obtained through careful
corrections of magnitudes across each plate. The overlapping
regions of neighbouring plates allowed us to obtain a uniform
instrumental system over the whole investigated region. The CCD
photometry for 1037 galaxies and 1085 stars in 92 fields allowed
one to establish the $r_F$ magnitudes for each galaxy in the
survey. The catalogue includes 5.5 million galaxies and is
complete up to $18^m.3$ (Ungruhe 1999).

\section{Analysis and result}

We selected a fragment of the ESO/SERC 598 field with Abell
cluster 2333. The A2333 cluster has 75 members, while in the field
from $20^h56^m.4$ to $21^h02^m.4$ (R.A.) and from $-18^\circ45'$
to $-19^\circ45'$ (Dec.) there are 1781 galaxies. In the first
stage of our investigation, we performed an analysis taking into
account galaxy positions only. The galaxy positions provide input
as seeds for the 2D Voronoi tessellation, and the Voronoi cell
around each galaxy is interpreted as the effective area that each
galaxy occupies in the projection plane. The result is shown in
Fig. 1. Taking the inverse of these areas will be the next step of
our procedure. In such a manner, the 2D local density of galaxies
is determined at each point containing a galaxy. This information
is then used to threshold and select galaxy members that live in
highly overdense regions, which we identify as clusters. We do so
by calculating the density contrast at each galaxy position as:
$\Delta = (\rho-<\rho>)/<\rho>=(<\sigma>-\sigma)/\sigma$ , where
$\sigma$ is the area of the Voronoi cells, and $<\sigma>$ is the
mean area of all cells.

\begin{figure}
\centering
  \includegraphics{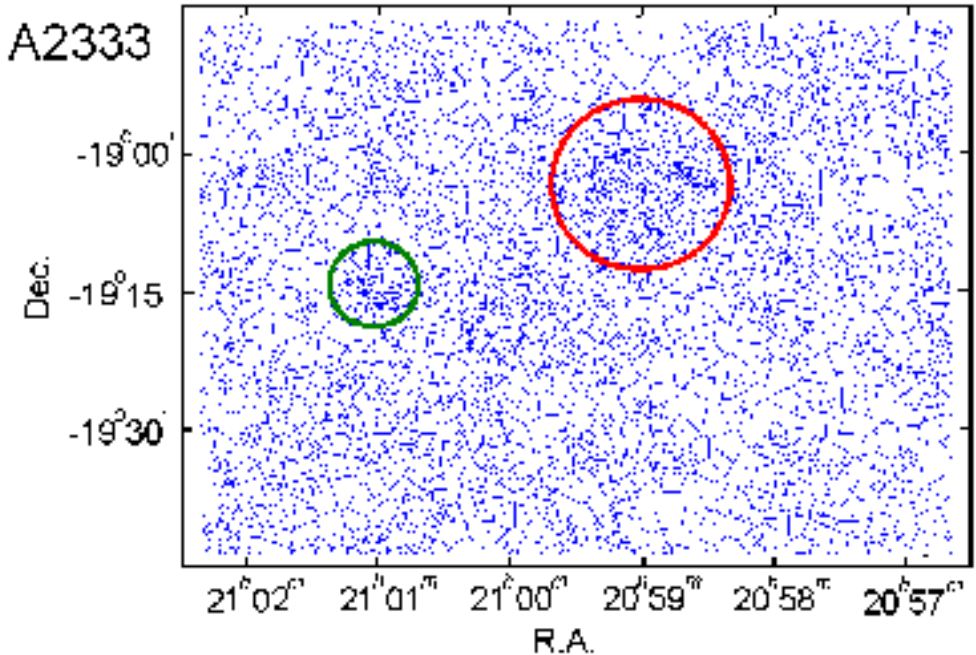}
  \caption{Region containing the A2333 after Voronoi tessellation. Each
Voronoi  cell encloses one galaxy. At  least one more dense region
is clearly seen.} \label{fig3}
\end{figure}

\section{Discussion and summary}
We show that the Voronoi tessellation allows us to find the
overdense regions. The main point of a further step in the search
of galaxy clusters is the determination of the contrast level.
This is crucial for finding real clusters. This can be done using
various methods. One of them consist in using the number
overdensity in respect to the average background. Another
possibility is to use the colour-magnitude relation combining the
data from the M\"{u}nster Red Sky Survey with those from the APM
catalogues. The determination of photometric redshifts can be
considered as well. At present, we are able to conclude that the
applied analysis allows us to find correctly the overdense regions
in two-dimensional data.

\begin{acknowledgments}
We thank Dr R. Ungruhe for sharing his data before their
publication. For this research we have used NASA's Astrophysics
Data System. PF has been partially supported by the Pedagogical
University grant BS 052.
\end{acknowledgments}


\begin{thebibliography}{}


\bibitem [Icke \& van de Weygaert (1987)]{Icke87}{Icke, V., \& van de Weygaert, R.} 1987, A\&A, 184, 16
\bibitem [Kepner, Fan, Bahcall, Gunn, Lupton, Xu (1999)]{Kepner99} {Kepner J., Fan X., Bahcall N., Gunn J., Lupton R., Xu G.} 1999, ApJ, 517, 78
\bibitem [Kim, Kepner, Postman, Strauss, Bahcall (2002)]{Kim02} {Kim R.S.J., Kepner J.V., Postman M., Strauss, M.A., Bahcall, N.A. et al.} 2002, AJ, 123, 20
\bibitem [Postman, Lubin, Gunn, Oke, Hoessel, Schneider \& Christensen
(1996)]{Postman96} {Postman M., Lubin, L., Gunn, J. E., Oke, J.
B., Hoessel, J. G., Schneider, D. P., \& Christensen, J. A.} 1996,
AJ, 111, 615
\bibitem [Ramella, Nonino, Boschin \& Fadda (1999)]{Ramella99} {Ramella, M., Nonino, M., Boschin, W., \& Fadda, D.} 1999, in ASP Conf. Ser. 176, Observational Cosmology: The Development of Galaxy Systems, ed. G. Giuricin, M. Mezzetti, \& P. Salucci (San Francisco: ASP), 108
\bibitem [Ramella, Boschin, Fadda, Nonino (2001)]{Ramella01} {Ramella M., Boschin W., Fadda D., Nonino M.} 2001, astro-ph/0101411 = A\&A 368, 776
\bibitem [Ungruhe (1999)]{Ungruhe99} {Ungruhe} 1999, Ph.D. thesis, Astron. Inst. Univ. M\"{u}nster, Germany.
\bibitem [Zaninetti (1989)]{Zaninetti89} {Zaninetti, L.} 1989, A\&A, 224, 345

\end{thebibliography}
\end{document}